\documentstyle[epsfig,twoside,rotating]{aa}

\newcommand{\degree}{^\circ}
\newcommand{\gsct}{\gamma \ {\rm Sct} }
\newcommand{\gnor}{\gamma \ {\rm Nor} }
\newcommand{\bsct}{\beta \ {\rm Sct} }
\newcommand{\tmus}{\theta \ {\rm Mus} }

\begin{document}
\title{Observation of Microlensing towards the
Galactic Spiral Arms. 
{\sc EROS II} 3 year survey
\thanks{This work is based on observations made at the
European Southern Observatory, La Silla, Chile.}
}
\author{{\sc EROS} Collaboration}

\author{
F.~Derue\inst{1}, 
C.~Afonso\inst{2},
C.~Alard\inst{3},
J.-N.~Albert\inst{1},
J.~Andersen\inst{4},
R.~Ansari\inst{1}, 
\'E.~Aubourg\inst{2}, 
P.~Bareyre\inst{5}, 
F.~Bauer\inst{2},
J.-P.~Beaulieu\inst{6},
G.~Blanc\inst{2},
A.~Bouquet\inst{5},
S.~Char\inst{7}\thanks{deceased},
X.~Charlot\inst{2},
F.~Couchot\inst{1}, 
C.~Coutures\inst{2}, 
R.~Ferlet\inst{6},
J.-F.~Glicenstein\inst{2},
B.~Goldman\inst{2},
A.~Gould\inst{8}\thanks{Alfred P. Sloan Foundation Fellow},
D.~Graff\inst{2,9},
M.~Gros\inst{2}, 
J.~Ha\"{\i}ssinski\inst{1}, 
J.-C.~Hamilton\inst{5},
D.~Hardin\inst{2},
J.~de Kat\inst{2},
A.~Kim\inst{5},
T.~Lasserre\inst{2},
\'E.~Lesquoy\inst{2},
C.~Loup\inst{6},
C.~Magneville \inst{2}, 
B.~Mansoux\inst{1}, 
J.-B.~Marquette\inst{6},
\'E.~Maurice\inst{10}, 
A.~Milsztajn \inst{2},  
M.~Moniez\inst{1},
N.~Palanque-Delabrouille\inst{2},
O.~Perdereau\inst{1},
L.~Pr\'evot\inst{10},
N.~Regnault\inst{1},
J.~Rich\inst{2}, 
M.~Spiro\inst{2},
A.~Vidal-Madjar\inst{6},
L.~Vigroux\inst{2},
S.~Zylberajch\inst{2}
\\   \indent   \indent
The {\sc EROS} collaboration\\
}
\institute{
Laboratoire de l'Acc\'{e}l\'{e}rateur Lin\'{e}aire,
{\sc IN2P3-CNRS}, Universit\'e de Paris-Sud, B.P. 34, 91898 Orsay Cedex, France
\and
{\sc CEA}, {\sc DSM}, {\sc DAPNIA},
Centre d'\'Etudes de Saclay, 91191 Gif-sur-Yvette Cedex, France
\and
{\sc DASGAL}, {\sc INSU-CNRS}, 77 avenue de l'Observatoire, 75014 Paris, France
\and
Astronomical Observatory, Copenhagen University, Juliane Maries Vej 30, 
2100 Copenhagen, Denmark
\and
Coll\`ege de France, {\sc LPCC}, {\sc IN2P3-CNRS}, 
11 place Marcellin Berthelot, 75231 Paris Cedex, France
\and
Institut d'Astrophysique de Paris, {\sc INSU-CNRS},
98~bis Boulevard Arago, 75014 Paris, France
\and
Universidad de la Serena, Facultad de Ciencias, Departamento de Fisica,
Casilla 554, La Serena, Chile
\and
Department of Astronomy, Ohio State University, Columbus, Ohio 43210, U.S.A.
\and
Department of Physics, Ohio State University, Columbus, Ohio 43210, U.S.A.
\and
Observatoire de Marseille, {\sc INSU-CNRS},
2 place Le Verrier, 13248 Marseille Cedex 04, France
}

\offprints{R. Ansari: ansari@lal.in2p3.fr; \\
{\it see also our WWW server at  URL :} \\
{\tt http://www.lal.in2p3.fr/recherche/eros}}
\date{Received ??/??/1999, accepted }

\thesaurus{10.08.1;10.11.1;10.19.2;10.19.3;12.04.1;12.07.1}

\maketitle
  \markboth{F. Derue et al. : Observation of Microlensing towards the
Galactic Spiral Arms}{F. Derue et al. : 
Observation of Microlensing towards the
Galactic Spiral Arms}

\begin{abstract}
We present an  analysis of the light curves of
9.1 million stars observed during three seasons
by {\sc EROS} (Exp\'erience de Recherche
d'Objets Sombres), in the Galactic plane away from the bulge.
Seven stars exhibit luminosity variations
compatible with gravitational microlensing effects due to
unseen objects.
The corresponding optical depth, averaged over four directions,
is $\bar\tau = 0.45^{+0.24}_{-0.11} \times 10^{-6}$.
While this value is compatible with expectations from simple 
galactic models under reasonable assumptions 
on the target distances, we find an excess of events with short 
timescale towards the direction closest to the 
Galactic Centre.
\end{abstract}

\keywords{Galaxy: Bar -- Galaxy: kinematics and dynamics -- Galaxy: stellar content -- Galaxy: structure -- {\itshape (Cosmology:)} gravitational lensing 
               }

\section{Introduction}
Extensive photometric surveys, triggered by Paczy\'nski's 
suggestion (1986), have led to the observation of 
microlensing effects towards the Magellanic clouds 
(EROS, \cite{eroslmc}; MACHO, \cite{machlmc} ) and
the Galactic bulge (OGLE, \cite{oglcg}; MACHO, \cite{machobulbe}).

The few hundred events observed towards the Galactic Centre 
(\cite{oglcg}; \cite{MachoBulge-1an}) have strengthened  
the hypothesis of a barred structure.
The early suggestion of de Vaucouleurs (1964) 
that the Galaxy is barred is now supported by many other observations 
including photometric measurements 
(\cite{Dwek}), studies of gas (\cite{weiner}), stellar kinematics
(\cite{Zhao-1996a}) and star counts (\cite{stanek}). 
Nevertheless, the bar parameters (shape, size, mass ...) 
are not yet precisely known.

In order to improve our knowledge of the Galactic structure,
{\sc EROS} started a dedicated observation program towards 
the Galactic Spiral Arms (GSA) in 1996. 
Four regions of the Galactic plane located at large angles 
from the Galactic Centre are now being monitored to disentangle 
the disc, bar and halo contributions
to the optical depths.
Three events with long 
Einstein crossing time have already been published,
based on two year (1996-97) EROS observations
(\cite{deruea}, hereafter paper I).
Because of their long duration, they are more easily interpreted as 
lensing events due to disc objects, rather than to halo deflectors.
We present in this paper an analysis of the three-year data set (1996-1998). 

\section{Experimental setup and observations}
The telescope, camera and observations, as well as the operations and 
data reduction are described in paper I and references therein. 
Four different directions are being monitored in the Galactic plane, 
corresponding to a total of 29 fields
with high stellar densities, covering
a wide range of Galactic longitude.
The three year data set contains 9.1 million light curves : 2.1 towards 
$\bsct$, 1.8 towards $\gsct$, 3.0 towards $\gnor$  
and 2.2 towards $\tmus$.
The observations span the period between July 1996 and November 1998,
except for $\tmus$ which is being monitored only since January 1997. 
An average of 100 measurements per field were obtained
in each of the $R_{EROS}$ and $V_{EROS}$ bands,
which are close to Cousins I and Johnson R to $\pm 0.3$ magnitudes.
As indicated in paper I, the distance distribution of our 
source stars is not precisely known.
We adopt an average distance for source stars of $\sim 7$ {\rm kpc}
for the discussion presented in this paper.
\section{The search for lensed stars}
\subsection{Data analysis selection results}

The data analysis is similar to that of the 
first two years, except that no rejection criteria based 
on the colour-magnitude diagram were applied.
This was made possible by the longer time coverage, 
allowing a better rejection against variable stars. 
The first step of the event selection filter
requires the presence of a single bump, simultaneous in the 
two EROS colours. To reject ordinary variable stars, we use the 
combined $\chi^{2}_{ml-out}$ of the microlensing fit
from light curves in both colours, estimated outside the peak 
(i.e. restricted to periods where the fitted
magnification is lower than 10\%).
We retain high signal-to-noise ratio events by requiring a significant 
improvement of the microlensing fit (ml) over a constant flux fit (cst):
\begin{eqnarray*}
\Delta = \frac{\chi^{2}_{cst} - 
\chi^{2}_{ml}}{\chi^{2}_{ml} / N_{dof}} \frac{1}{\sqrt{2N_{dof}}} 
> \Delta _{min}\ ( = 15) 
\end{eqnarray*}
%
%
%
where $N_{dof}$ is the number of degrees of freedom.
\begin{figure}[h!]
\begin{center}
\begin{turn}{0}
\mbox{\epsfig{file=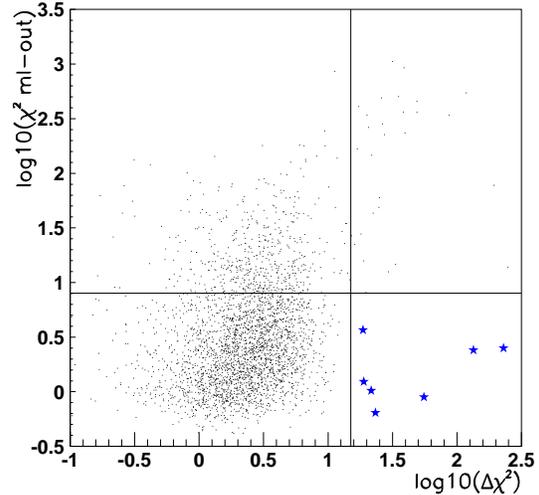,width=7.5cm}}
\end{turn}
\caption[]{Distribution of $\log_{10}(\chi^{2}_{ml-out})$ versus 
$\log_{10}(\Delta)$ for a subset of all light curves. 
The two lines correspond to the adopted cuts. 
Stars ($\star$) correspond to the seven candidates selected by our analysis.}
\label{analyse}
\end{center}
\end{figure}

Seven light curves satisfy all the requirements and are
labelled GSA1 to 7.
Fig. \ref{analyse} shows the distribution of $\log_{10}(\chi^{2}_{ml-out})$ 
versus $\log_{10}(\Delta)$ for lightcurves satisfying all the other 
selection criteria.
The seven candidates are located in a region of the diagram corresponding 
to lightcurves with a magnification well described by a microlensing fit 
and constant outside the peak.
The upper right side of the diagram is populated by variable stars, 
mostly red and bright.
\begin{table*}
\caption[]{Characteristics of the 7 microlensing candidates and
contribution to each direction's optical depth.}
\begin{center}
\begin{tabular}{llllllll}
\hline\noalign{\smallskip}
Candidate & GSA1 & GSA2 & GSA3 & GSA4	& GSA5	& GSA6 & GSA7 \\
\noalign{\smallskip}
\hline\noalign{\smallskip}
field & $\gsct$ & $\gnor$ & $\gnor$  & $\gsct$ & 
 $\gsct$ &  $\gsct$ &  $\gsct$  \\
$\alpha$(h:m:s)	eq.2000 & 18:29:09.0	& 16:11:50.2	& 16:16:26.7 & 18:32:26.0	& 18:32:12.0& 18:33:45.5& 18:35:12.4	\\
$\delta$(d:m:s) eq.2000	& -14:15:09& -52:56:49& -54:37:49 & -12:56:04& -12:55:16	& -14:41:41& -14:56:27 \\
$R_{EROS}$ - $V_{EROS}$	& 17.7 - 20.7		& 17.8 - 19.4 		& 17.5 - 18.6 & 17.1 - 17.9		& 17.9 - 19.9		& 17.2 - 18.5 & 17.5 - 18.7 \\
$\Delta t = R_E/V_T$ (days) & $73.5\pm 1.4$	& $98.3\pm 0.9$	& $70.0\pm 2.0$ & $23.9\pm 1.1$	& $59.0\pm 5.5$	& $37.9\pm 5.0$	& $6.20 \pm 0.50$ \\
Max. magnification	& $26.5\pm 0.6$	& $3.05\pm 0.02$	& $1.89\pm 0.01$ & $1.72\pm 0.02$	& $1.71\pm 0.03$	& $1.35\pm 0.02$& $2.70 \pm 0.30$	\\
contribution to $\tau$ ($\times 10^6$) & 0.51 & 0.15 & 0.12 & 0.30 & 0.44 & 0.35 & 0.22 \\
\noalign{\smallskip}
\hline
\end{tabular}
\end{center}
\label{caract}
\end{table*}
Table \ref{caract} contains the characteristics of the 7 candidates.
GSA1 \& 2 have been studied in detail in paper I 
leading to additional constraints on lens masses and distances.
None of the new candidates shows any noticeable 
deviation from standard microlensing curves. 
Finding charts and lightcurves can be found in \cite{deruea} 
(see also our WWW server).
\subsection{The analysis efficiency}
To determine the efficiency of each selection criterion, we
have applied them to Monte-Carlo generated light curves,
obtained from a representative sample of the observed
light curves on which we superimpose randomly generated
microlensing effects.
\begin{table}[h!]
\begin{center}
\caption[]{Detection efficiency (in \%) of the analysis as a 
function of the event timescale.
}
\begin{tabular}{|l|c|c|c|c|}
\hline
Event timescale          & \multicolumn{4}{|c|}{Direction}\\ 
 (days)& $ \bsct$  & $\gsct$ & $\gnor$ & $\tmus$\\
\hline
$\Delta t=6$   & 2.1  & 2.1 & 5.3  & 9.4  \\
$\Delta t=25$  & 6.0  & 5.1 & 12.5 & 14.5 \\
$\Delta t=40$  & 8.5  & 8.0 & 17.0 & 17.0 \\
$\Delta t=70$  & 11.5 & 10.5 & 25.0 & 18.7 \\
$\Delta t=100$ & 13.8 &10.5 & 30.0 & 18.5 \\ 
\hline
\end{tabular} 
\end{center}
\label{efficiency} 
\end{table}
The microlensing parameters are uniformly drawn in the
following intervals:
impact parameter expressed in units of the Einstein radius 
$u_0$~$\in$~$[0,2]$, maximum magnification time in a search period $T_{obs}$
starting 150 days before
the first observation and ending 150 days after the last 
observation,
and Einstein radius crossing time $\Delta t$~$\in$~$[1,250]$ days.
As in paper I, the analysis efficiency 
(or sampling efficiency) $\epsilon(\Delta t)$ 
reported in
Table 2 is relative to a set of unblended stars, normalised to $u_0 < 1$.
\section{Galaxy model, optical depth and event timescale}
\begin{figure}[h!]
\begin{center}
\mbox{\epsfig{file=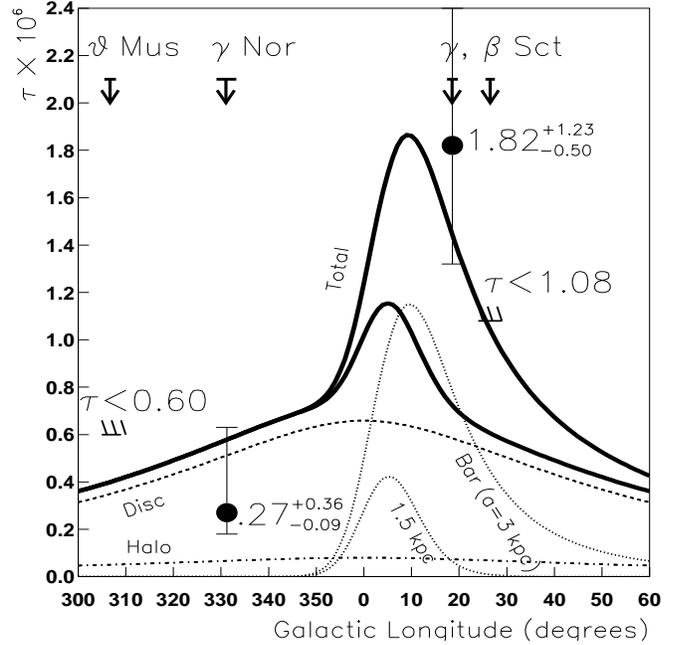,width=9.5cm,height=8.5cm,angle=0.}}
\caption[]{Expected optical depth ($\times 10^{6}$) up to $7\ \rm{kpc}$
for the different components of the Milky Way as a function of
the Galactic longitude at $b= -2.\hskip-2pt \degree 5$ for two values of 
the bar length parameter ($a$).
The 4 directions towards the spiral arms are indicated by vertical arrows
whose horizontal dash shows the field extension in longitude.}
\label{optvsl}
\end{center}
\end{figure}

%
%
We have computed the expected optical depth (probability of observing 
a magnification larger than 1.34 for a pointlike source) using 
a three component model for the deflectors~:
a bulge, described by a barlike triaxial distribution, a thin disc, 
and a standard isotropic and isothermal halo (same as model 1 in paper I).
The expected optical depth averaged over the four directions is 
$0.60 \times 10^{-6}$ for this model.
Fig. \ref{optvsl} shows the optical depth up to $7\ {\rm kpc}$
as a function of Galactic longitude for two values of the bar  
semi-major axis ($a=1.5\ {\rm kpc}$ as in paper I 
and $a=3\ {\rm kpc}$),
at the average latitude of our fields $b=-2.5\degree$.
Note that at this Galactic latitude, the bulge and disc contributions are
already reduced by a factor $\simeq 2$ with respect
to zero latitude; moreover, the bulge contribution also 
changes dramatically with the distance to the target (here 
assumed to be at 7 kpc, while the Galactic centre is at 8.5 kpc).


Assuming a standard halo completely made of 
compact objects would lead to
a halo contribution of less than 10\% to the GSA optical depth. 
Moreover the {\sc EROS} measurements towards the {\sc LMC} \linebreak
(\cite{Eros1LMC}; \cite{ErosMacho}; \cite{lasserre}) 
and the {\sc SMC} (\cite{afonso}) suggest that no more than 40\%  of 
this halo can be made of MACHOs lighter than $0.5 M_\odot$ . 
Thus, for the sake of clarity, we will neglect the halo 
contribution in the following discussion.

\begin{figure}[h!]
\begin{center}
\mbox{\epsfig{file=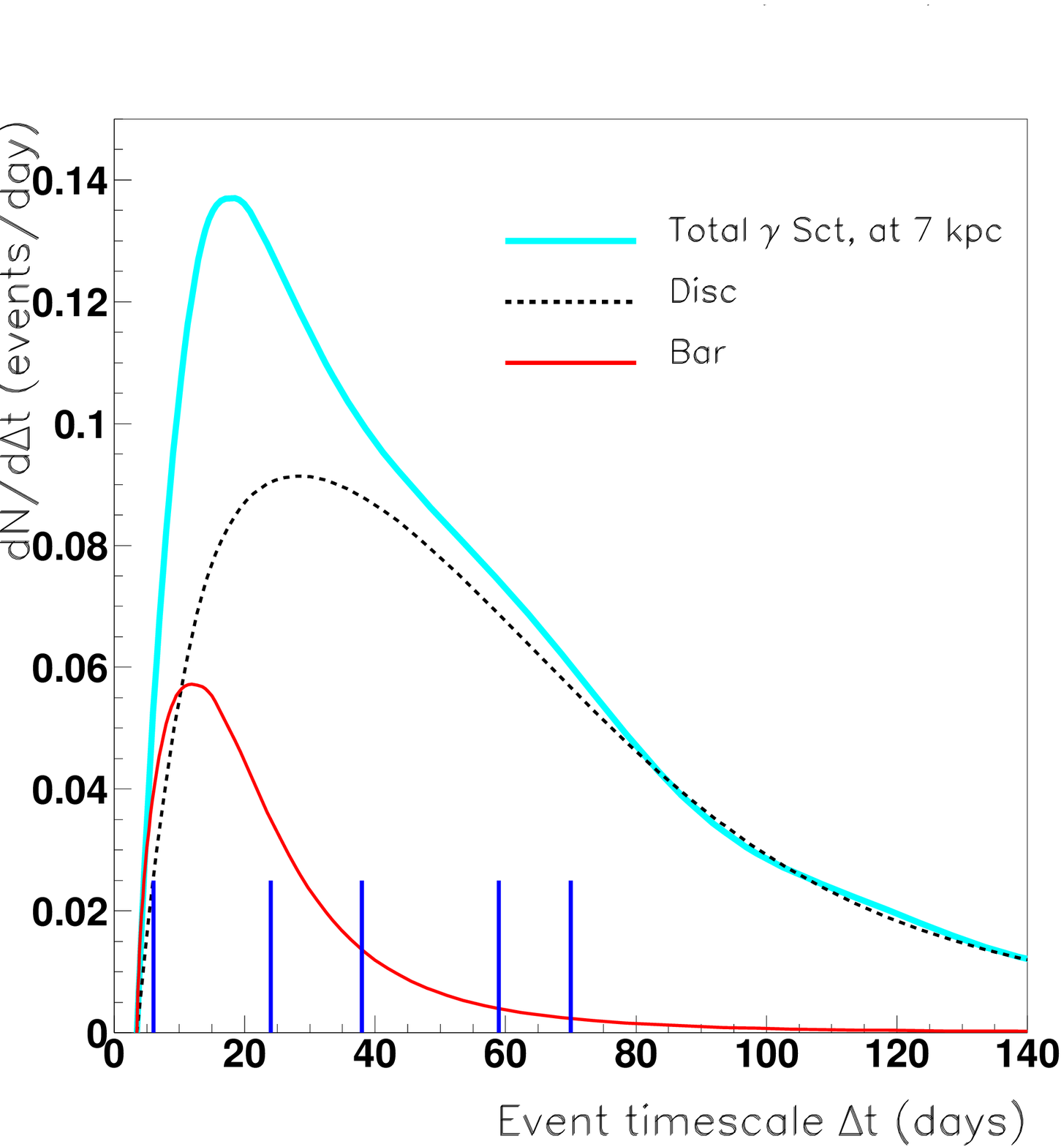,width=8.5cm,height=7.0cm,angle=0}}
\caption[]{Expected rate versus event timescale towards $\gsct$. 
Detection efficiencies $\epsilon(\Delta t)$ are taken into account.}
\label{duration}
\end{center}
\end{figure}

Fig. \ref{duration} shows the expected event duration 
distribution towards $\gsct$. The durations for the 
five events observed toward  $\gsct$ are also indicated.
The predicted distribution is obtained using the kinematic characteristics 
and mass functions given in paper I.
As the disc lenses have a low velocity relative to the line of sight,
disc-disc events have longer timescales ($\sim$~$60$ days) than 
bar-disc events ($\sim$~$20$ days).

\section{Optical depth estimation}
For a given target, an estimate of the optical depth 
or a limit (when $N_{evt} = 0$) can be computed 
using the expression:
\begin{eqnarray*}
\tau = \frac{1}{N_{obs}T_{obs}}\frac{\pi}{2}\sum_{events}
\frac{\Delta t}{\epsilon (\Delta t )} \ , 
\label{tau}
\end{eqnarray*}
where $N_{obs}$ is the number of monitored stars in
the target and $T_{obs}$ the duration of the search period
(1170 days for this 3 year analysis, except 990 days towards $\tmus$).

We compute an average optical depth $\bar\tau$ by a weighted mean 
($w = N_{obs} \times T_{obs}$) over the four directions.
We find the value $\bar\tau = 0.45^{+0.24}_{-0.11} \times 10^{-6}$, 
in agreement with expectations.
%
%
 
In Fig. \ref{optvsl} we report for each target the measured optical depth. 
The quoted errors include only Poisson fluctuations and indicate the bayesian 
68\% confidence intervals.
In the case of $\bsct$ and $\tmus$, where no events were 
observed, we have computed a 95\% confidence level upper limit,
assuming a mean event duration of 50 days~: 
$\tau(\tmus ) <0.60 \times 10^{-6}$ and 
$\tau(\bsct ) <1.08 \times 10^{-6}$.
 
The two targets $\gsct$ and $\gnor$ are located at nearly symmetric
longitudes with respect to the Galactic Centre.
Yet, we find an optical depth toward $\gsct$ ($\simeq 1.82 \times 10^{-6}$)
significantly higher (at more than $2\sigma$)
than toward $\gnor$ ($\simeq 0.27 \times 10^{-6}$).
In addition, the average measured event timescale toward $\gsct$
is 40 days, half of that observed for $\gnor$.
These features suggest a significant contribution from 
the bar toward $\gsct$.
Although an asymmetry is expected from the bar contribution in
model I, our observations indicate a larger difference in
optical depths. Indeed we find a deficit in the event rate
toward $\gnor$ and an excess toward $\gsct$, compared
to our model's predictions, as can be seen in
figure 2.

Provided that this is not due to a statistical fluctuation, at least two
simple hypotheses could explain the observed asymmetry: \\
- An increase in the bar length parameter enhances 
the asymmetric contribution to the optical depth.
Changing this parameter from $a=1.5$ to $a=3$ kpc leads to 
an optical depth toward $\gsct$ $\tau(\gsct ) = 1.40 \times 10^{-6}$. \\
- The optical depth is very sensitive to the poorly known distance 
distribution of the monitored source stars, which depends on 
the star number density and the extinction along the line of sight. 
For example, changing the 
$\gsct$ source star distances from 7 kpc to 9 kpc (resp 11 kpc) 
increases the expected optical depth from $0.75 \times 10^{-6}$ to 
$1.3 \times 10^{-6}$ (resp. $1.93 \times 10^{-6}$). However, 
this hypothesis alone cannot account for the shorter 
event durations observed toward $\gsct$.

\section{Conclusion}
We have searched for microlensing events with durations
ranging from a few days to a few months in four Galactic disc
zones lying at $18^\circ$ to $55^\circ$ from the Galactic Centre.
We find seven events that can be interpreted as microlensing effects due to
massive compact objects. The estimated average optical depth is 
compatible with expectations from simple Galactic models.
However, we observe variations of the event rate 
with the Galactic longitude which differ from these models predictions.
Additional data is needed to confirm the reported discrepancy
and to shed light on its interpretation.

\begin{acknowledgements}
We are grateful to D. Lacroix and the technical staff at the Observatoire
de Haute Provence and to A. Baranne for their help in 
refurbishing the {\sc MARLY}
telescope and remounting it in La Silla. We are also grateful for
the support given to our project by the technical staff at ESO, La Silla.
We thank J.F. Lecointe for assistance with the online computing.
We wish to thank also C. Nitschelm for his contribution to the data taking.
\end{acknowledgements}

\end{document}